\documentclass[fleqn,usenatbib,useAMS]{mnras}

\usepackage{graphicx}	% Including figure files
\usepackage{amsmath}	% Advanced maths commands
\usepackage{amssymb}	% Extra maths symbols
\usepackage{multicol}        % Multi-column entries in tables
\usepackage{bm}		% Bold maths symbols, including upright Greek
\usepackage{pdflscape}	% Landscape pages
\usepackage{ulem}
\newcommand{\degree}{\ensuremath{^\circ}}

\usepackage[T1]{fontenc}
\usepackage{ae,aecompl}

\usepackage{newtxtext,newtxmath}
\date{Last updated 2020 June 10; in original form 2013 September 5}

\pubyear{2022}

\title[ Pulsar wind nebulae of runaway massive stars ]{ Pulsar wind nebulae of runaway massive stars }

\author[Meyer and Meliani]
       {Meyer, D. M.-A.$^1$ and Meliani, Z.$^2$ \\ 
       $1$ Institut für Physik und Astronomie, Universität Potsdam, Karl-Liebknecht-Strasse 24/25, D-14476 Potsdam, Germany,\\
       E-mail: dmameyer.astro@gmail.com  \\ 
       $2$ Laboratoire Univers et Théories, Observatoire de Paris, Université PSL, Université de Paris, CNRS, F-92190 Meudon, France\\
       E-mail: zakaria.meliani@observatoiredeparis.psl.eu  \\
       }

\begin{document}
\label{firstpage}
\pagerange{\pageref{firstpage}--\pageref{lastpage}}
\maketitle

\begin{abstract}
\textcolor{black}{
A significant fraction of massive stars move at speed through the interstellar medium of galaxies. 
After their death as core-collapse supernovae, a possible final evolutionary state is that of a fast-rotating magnetised neutron star -- shaping its circumstellar medium into a pulsar wind nebula. Understanding the properties of pulsar wind nebulae requires knowledge of the evolutionary history of their massive progenitors. Using two-dimensional magneto-hydrodynamical simulations, we demonstrate that, in the context of a runaway high-mass 
red-supergiant supernova progenitor, the morphology of its subsequent pulsar wind nebula is strongly affected by the wind of the defunct progenitor star pre-shaping the stellar surroundings throughout its entire past life. 
In particular, pulsar wind nebulae of obscured runaway massive stars harbour asymmetries function of the morphology of the progenitor’s wind-blown cavity, inducing projected asymmetric up-down synchrotron emission.
}
\end{abstract}

\begin{keywords}
methods: MHD -- stars: evolution -- stars: massive -- pulsars: general -- ISM: supernova remnants.
\end{keywords}

%%%%%%%%%%%%%%%%%%%%%%%%%%%%%%%%%%%%%%%%%%%%%%%%%%%%%%%%%%%%%%%%%%%%%%%%%%%%%%%%%%%%%%%%%%%
%%%%%%%%%%%%%%%%%%%%%%%%%%%%%%%%%%%%%%%%%%%%%%%%%%%%%%%%%%%%%%%%%%%%%%%%%%%%%%%%%%%%%%%%%%%
%%%%%%%%%%%%%%%%%%%%%%%%%%%%%%%%%%%%%%%%%%%%%%%%%%%%%%%%%%%%%%%%%%%%%%%%%%%%%%%%%%%%%%%%%%%

\section{introduction}
%%%%%%%%%%%%%%%%%%%%%%%%
% Observations of runaway star->SNR
%%%%%%%%%%%%%%%%%%%%%%%%
About $10\%$$-$$25\%$ of OB-type massive stars are runaway
objects~\citep{Blaauw_1961BAN....15..265B,Schoettleretal_2022MNRAS.510.3178S}  with peculiar 
supersonic speeds with respect to their interstellar medium (ISM). 
\textcolor{black}{
Two main mechanisms are proposed to explain the production of such runaway stars,  
\textcolor{black}{namely} binary-supernova ejection and dynamical ejection from the stellar cluster~\citep{Hoogerwerf_2000ApJ...544L.133H} 
}
and their trajectory can be traced back to their parent starburst 
regions~\citep{Schoettler_2019MNRAS.487.4615S}. 
These massive stars die as core-collapse supernovae releasing most of the remaining stellar mass with high speed into the surrounding circumstellar medium (CSM) before expanding into the ISM.

%%%%%%%%%%%%%%%%%%%%
% SNR-PWN
%%%%%%%%%%%%%%%%%%%%
%\textcolor{black}{
%The supernova forward shock sweeps-up, compresses and heats its surroundings, which induces a deceleration of its 
%associated reverse shock. [OK here ?, sounds weird. ].
%} 
%
Often, at the location of the supernova explosion, remains a highly-magnetised and fast-rotating neutron star (pulsar). The pulsar generates a powerful wind with a kinetic luminosity that can reach $\dot{E} \sim 10^{39}\, \rm erg\, \rm s^{-1} $ inducing a growing pulsar wind nebula (PWN). This PWN evolves first into the freely-expanding supernova ejecta before reaching the supernova reverse shock~\citep{Blondin_etal2001ApJ...563..806B}. After it interacts with the swept-up CSM and, at a later time, with the unshocked ISM.
%}
%\sout{unperturbed ISM.} 

%%%%%%%%%%%%%%%%%%%%%%%%
% models done
%%%%%%%%%%%%%%%%%%%%%%%%

Previous investigations on PWN of moving pulsar concentrate on the birth kick pulsar. It occurs when massive stars collapse~\citep{deVries_etal2021ApJ...908...50D}. The resulting pulsar propagates through the supernova ejecta~\citep{Slane_etal2018ApJ...865...86S}. 
On timescales of a few $\rm kyr$, the fast-moving pulsar escapes the supernova remnant (SNR) and travels in the uniform and cold ISM~\citep{Barkov_etal2019MNRAS.485.2041B, Bucciantini_etal2020JPhCS1623a2002B}.
The resulting PWN morphology is strongly deformed and develops an extended tail. Because of numerical and physical difficulties associated with the CSM modelling, the simulations of PWN are mainly performed in classical hydrodynamics and deal only with moving pulsars inside supernova ejecta~\citep{Temim_etal2015ApJ...808..100T,2022arXiv220501798T}. Relativistic simulations have been performed for pulsars moving in the ISM~\citep{Barkov_etal2019MNRAS.485.2041B, Bucciantini_etal2020JPhCS1623a2002B}, while others also deal with static pulsar wind expanding into the supernova ejecta in 1D and 
2D~\citep{Blondin_etal2001ApJ...563..806B,vanderswaluw_aa_397_2003,vanderswaluw_aa_404_2003,vanderswaluw_aa_420_2004,Slane_etal2018ApJ...865...86S}.
However, these works did not account for all the evolution phases of the stellar progenitor of the runaway pulsar.

%%%%%%%%%%%%%%%%%%%%%%%%
% What new with our model
%%%%%%%%%%%%%%%%%%%%%%%%
For pulsars from massive runaway stars, their wind mainly expands in the supernova remnant (SNR) constituted of supernova ejecta and CSM materials. Thus, shocks and instabilities, as well as the overall PWN morphology will be affected by these structures. Since SNRs are strongly affected by the CSM distribution~\citep{vanmarle_aa_537_2012}, the PWNs and the properties of the various shocks therein should be, in their turn, a function of their surrounding CSM.
Most common runaway massive stars shape the CSM with a large amount of red supergiant material as dense asymmetric  stellar wind bow shocks~\citep{vanmarle_aa_537_2012,henney_mnras_486_2019,henney_486_mnras_2019,henney_mnras_489_2019,meyer_mnras_493_2020,herbst_sstv_2022}, it should not be omitted in the modelling of their PWNs.

%%%%%%%%%%%%%%%%%%%%%%%%
% short description of the paper
%%%%%%%%%%%%%%%%%%%%%%%%
%
Motivated by the above arguments, we numerically investigate the shaping of PWN of fast-moving massive core-collapse supernova progenitor. We focus on the particular effects of the CSM generated by the wind-ISM interaction of the progenitor star. 
This study is organised as follows. First, we present the numerical methods used to model the pulsar wind nebula of a runaway massive progenitor star in Section~\ref{model}.  The outcomes of the simulations are presented in Section~\ref{results}. We discuss our results and draw our conclusions in Section~\ref{discussion}.

\begin{figure*}
        \centering
        \includegraphics[width=0.865\textwidth]{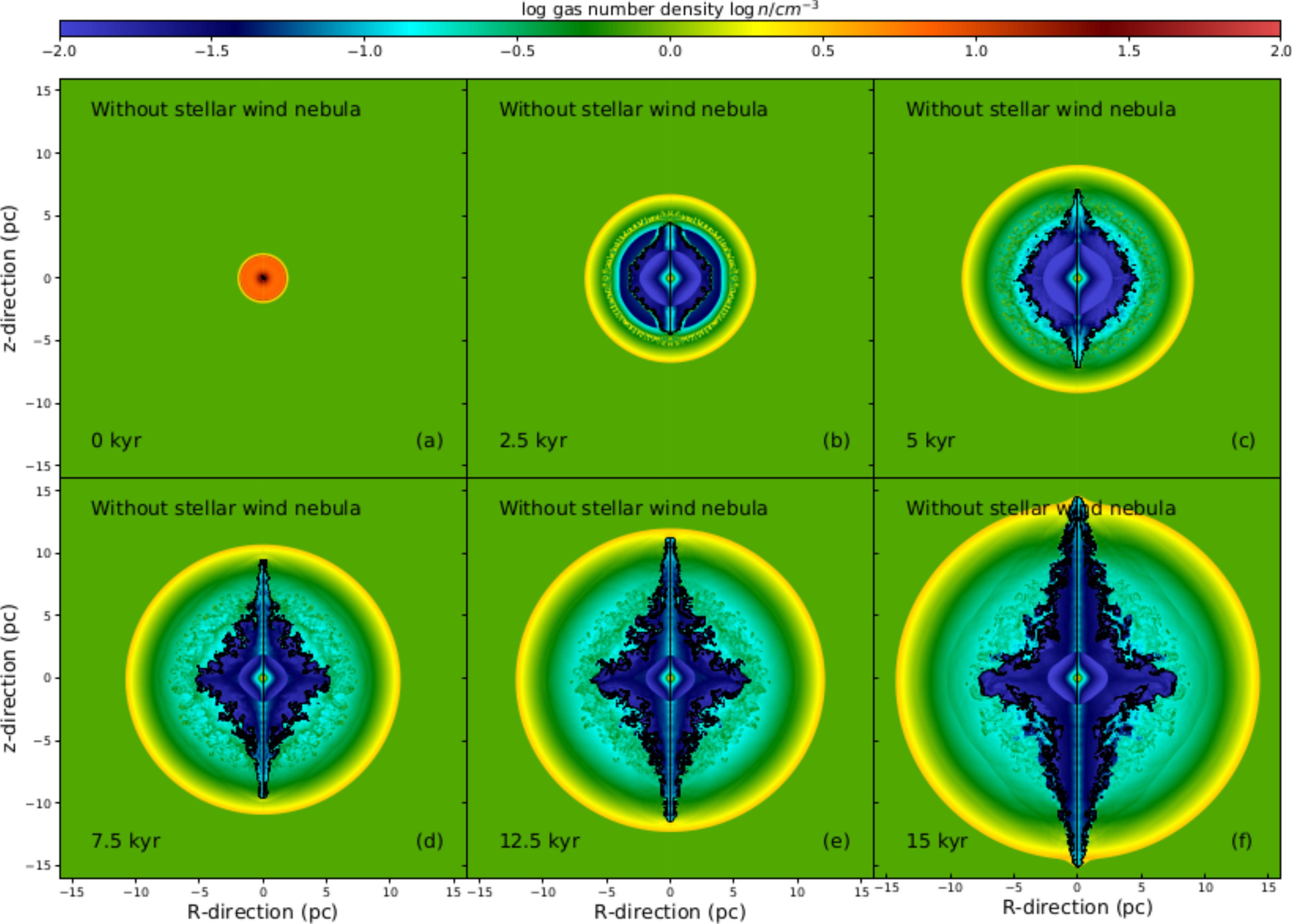}  \\        
        \caption{
        Time evolution of the 
        number density field (in ${\rm cm^{-3}}$) in the simulation of the nebula generated by a 
        pulsar wind injected into the expanding ejecta of a core-collapse supernova. 
        The black contour marks the region of the nebula made of $50\%$ of pulsar wind 
        material in number density. 
        Time is measured starting from the onset of the pulsar wind. 
        }
        \label{fig:plot_no_csm}  
\end{figure*}

\begin{figure*}
        \centering
        \includegraphics[width=0.865\textwidth]{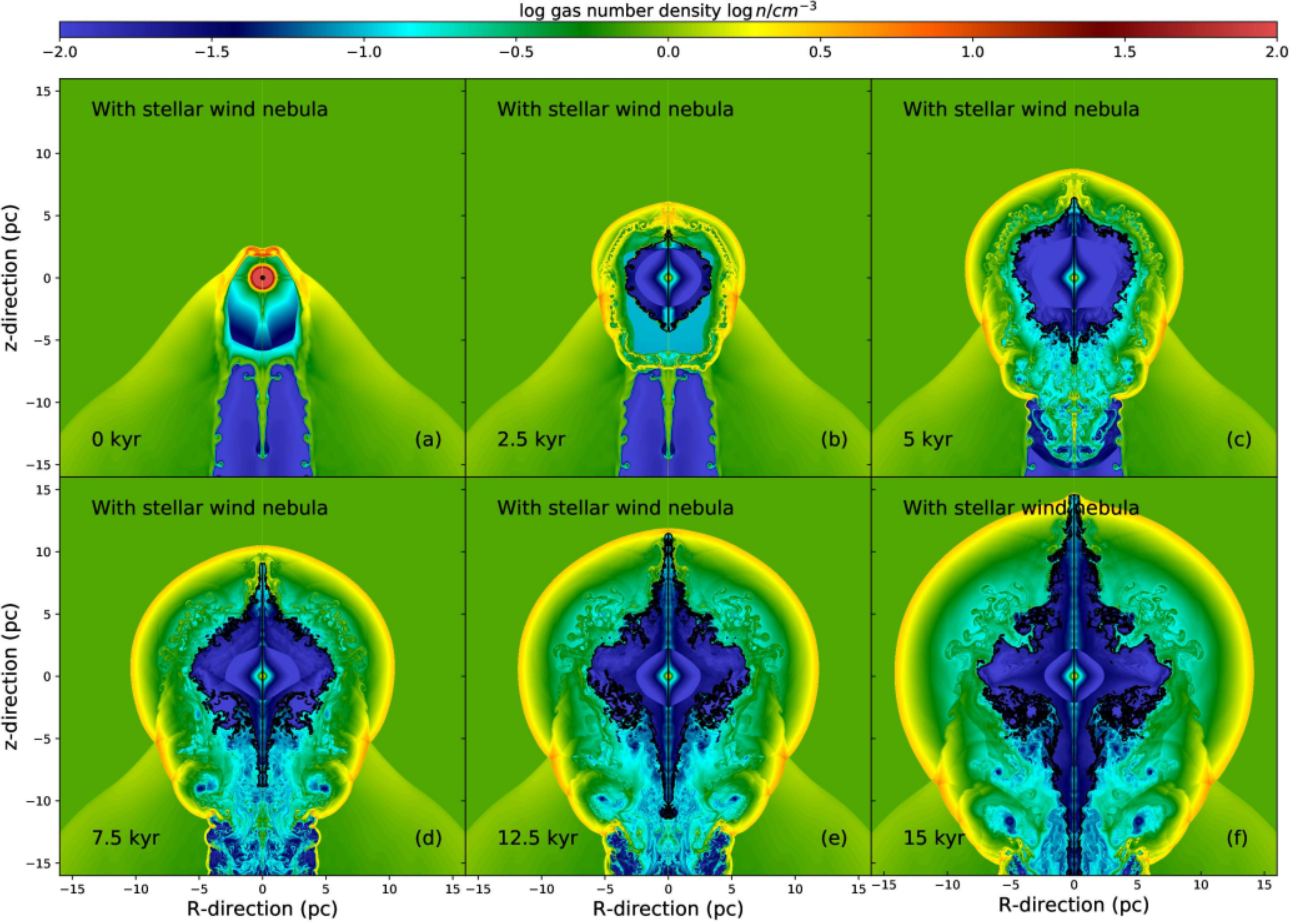}  \\        
        \caption{
        As for Fig.~\ref{fig:plot_no_csm}, with the presence of the circumstellar medium 
        (CSM) generated by the wind-ISM interaction of the $20\, \rm M_{\odot}$ 
        massive runaway progenitor. 
        }
        \label{fig:plot_csm}  
\end{figure*}

%%%%%%%%%%%%%%%%%%%%%%%%
% 20 solar mass star evolution
%%%%%%%%%%%%%%%%%%%%%%%%
\section{Method}\label{model}

%%%%%%%%%%%%%%
% stellar wind model 
%%%%%%%%%%%%%%
This study focuses on the entire evolution of the circumstellar medium of a runaway 
$20\, \rm M_{\odot}$ star at Galactic metallicity. The stellar surface properties 
such as wind velocity and mass-loss rate histories are taken from \textcolor{black}{a 
model of the {\sc geneva} evolutionary tracks library calculated without rotational 
mixing}~\citep{eldridge_mnras_367_2006,ekstroem_aa_537_2012}.

%%%%%%%%%%%%%%
% ISM characteristic 
%%%%%%%%%%%%%%
%
%The high temperature of the interstellar medium surrounding the massive star result from the 
%ionizing massive star radiation where the Strömgren sphere has radius 4.3 pc \citep{mackey_mnras_2013}.
%
%%%%%%%%%%%%%%
% SNR characteristics
%%%%%%%%%%%%%%
We model the stellar surroundings from the zero-age main-sequence to the pre-supernova phase, through 
its red supergiant phase. When the star achieves its evolution, we inject therein supernova core-collapse 
ejecta made of mass $M_{\rm ej}=6.96\, \rm M_{\odot}$ and energy 
$E_{\rm ej}=10^{51} \textrm{erg}$, using a density power-law 
profile~\citep{whalen_apj_682_2008,meyer_mnras_493_2020}. 
%
%
%
%%%%%%%%%%%%%%%%%%%%%%%%
% SNR and PWN
%%%%%%%%%%%%%%%%%%%%%%%%
%
%
%%%%%%%%%%%%%%
% PWN characteristics 
%%%%%%%%%%%%%%
The pulsar wind is modelled according to~\citet{komissarov_mnras_349_2004}, using a 
wind power $\dot{E}_{\rm o}=10^{38}\, \rm erg\, \rm s^{-1}$, a  wind velocity $10^{-2}c$ with $c$ the speed 
of light in vacuum, a pulsar spin $P_{\rm o}=0.3\, \rm s$, a spin period variation 
$\dot{P}_{\rm o}=10^{-17}\, \rm s\, \rm s^{-1}$ and a wind magnetisation parameter 
$\sigma=10^{-3}$~\citep{2017hsn_book_2159S}. The pulsar loses energy according to 
an initial spin-down time scale $\tau_{\rm o}=P_{\rm o}/((N-1)\dot{P}_{\rm o})$, 
where the braking index $N=3$ stands for magnetic dipole spin-down. 
The energy diminution equations therefore reads 
$\dot{E}(t)=\dot{E}_{\rm o}(1+t/\tau_{\rm o})^{\alpha}$ with 
$\alpha=-(N+1)/(N-1)$, see~\citet{pacini_apj_186_1973}. 
\textcolor{black}{
In this first paper,  we run 2.5D simulations. Therefor, it is assumed that the 
spin axis and the direction of motion of the star coincide with the axis of 
symmetry of the cylindrical coordinate system.
}
%
%
%%%%%%%%%%%%%%%%%%%%%%%%
% Why we can use classical wind for pulsar
%%%%%%%%%%%%%%%%%%%%%%%%

The massive star moves through the ISM of the Milky Way's galactic plane of initial gas density 
$0.79\, \rm cm^{-3}$ and temperature of $8000\, \rm K$ \citep{wolfire_apj_587_2003, meyer_2014bb}.
\textcolor{black}{
We adopt an undisturbed ISM magnetic field parallel to direction of motion of the star with a strength of $7\ \mu \rm G$ which corresponds to a Alfv\' en  
speed of $ v_{\rm A}= 17.2 \, \rm km\ \rm s^{-1}$~\citep{meyer_mnras_464_2017}. }
%
%\setout{Our 2.5 cylindrical coordinate system imposes us that the magnetic field has to be taken as parallel to the direction of stellar motion and to that of the pulsar spin axis. }

%%%%%%%%%%%%%%%%%%%%%%%%
% Equation in use and numerical method
%%%%%%%%%%%%%%%%%%%%%%%%
The $2.5$D numerical magneto-hydrodynamical simulations 
are performed with the code {\sc pluto}~\citep{mignone_apj_170_2007,migmone_apjs_198_2012}. We make use of the  
Godunov-type numerical scheme HLL Riemann solver combined with the PPM limiter, 
together with a third-order Rung-Kutta time-marching algorithm controlled by the Courant-Friedrich-Levi 
number. As in~\citet{vanderswaluw_aa_397_2003}, we set the polytropic index to $5/3$ 
and optically-thin radiative cooling physics is used for all circumstellar evolutionary 
phases except starting from the onset of the pulsar wind. 
More precisely, we make use of the optically-thin cooling and heating curves for fully ionized 
medium~\citep{wiersma_mnras_393_2009} of solar abundance~\citep{asplund_araa_47_2009} 
presented in~\citet{meyer_2014bb}.
%
%
%%%%%%%%%%%
% Grid 
%%%%%%%%%%%
Calculations are conducted following a mapping 
strategy~\citep{vanmarle_584_aa_2015,meyer_mnras_493_2020,meyer_mnras_502_2021} 
in which the progenitor's CSM is first calculated, 
before supernova ejecta and eventually pulsar wind are injected into it. 
The models are conducted using a cylindrical coordinate system ($R,z$) that is 
mapped with a uniform mesh $[0,150 ]\times[-50,50]$ of spatial 
resolution $1.2\times 10^{-2}\, \rm pc\, \rm cell^{-1}$ for the CSM and a mesh 
$[0,20]\times[-20,-20]$ of resolution 
$6.7\times10^{-3}\, \rm pc\, \rm cell^{-1}$ for the PWN. The central 
sphere in which winds are imposed is of radius 
$20\, \rm cells$ ($0.24\, \rm pc$ during the stellar wind phase and $0.134 \,\rm pc$ 
throughout the pulsar phase).

%%%%%%%%%%%%%%%%%%%%%%%
%Desciption of methode in use
%%%%%%%%%%%%%%%%%%%%%%%
Two simulations are performed with a runaway star of space velocity $v_{\star}=40\, \rm km\, \rm s^{-1}$ 
(Mach number $\mathcal{M}\sim 4$). The first one considers the pulsar wind expanding into supernova 
ejecta only, while the second one also accounts for the progenitor's CSM.

\begin{figure*}
        \centering
        \includegraphics[width=0.865\textwidth]{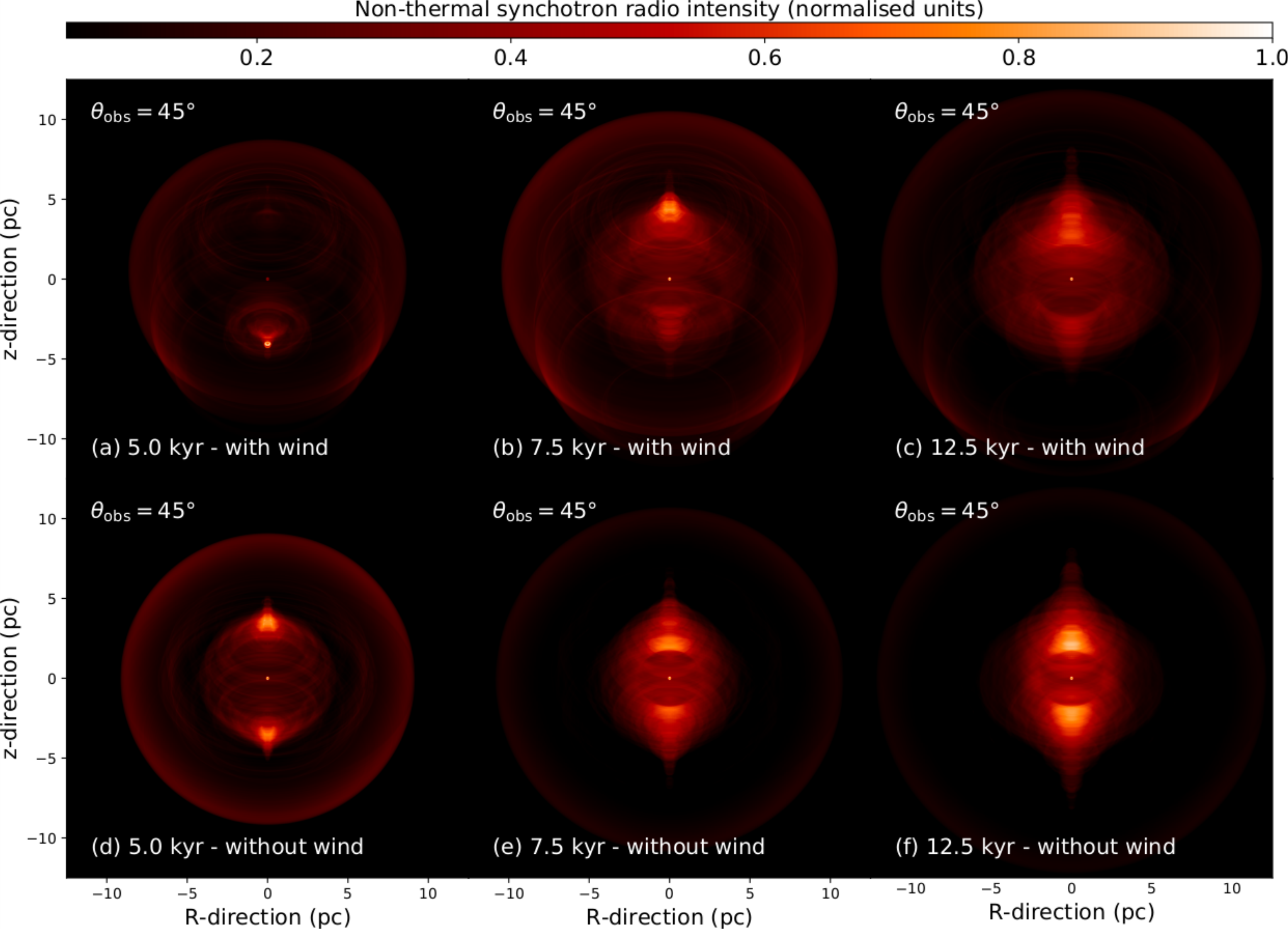}  \\        
        \caption{
        Selected time sequence evolution of synchrotron emission map 
        assuming an angle $\theta_{\rm obs}=45\degree$ between the observer's line-of-sight 
        and the axis of symmetry of the nebula. 
        The images compare models with (top) and without (bottom) the stellar wind 
        of the massive progenitor. 
        }
        \label{fig:plot_synchrotron_1}  
\end{figure*}

\section{Results}
\label{results}

%%%%%%%%%%%%%%%%%%%%%%%%%%%%%%%%%%%%%%%%%%%%%%%
\subsection{Model without wind-ISM interaction}
\label{result_1}
%%%%%%%%%%%%%%%%%%%%%%%%%%%%%%%%%%%%%%%%%%%%%%%
For the sake of comparison, we first model a PWN accounting for core-collapse supernova ejecta but neglecting the CSM, 
the entire system being in supersonic motion in the ambient medium. The time sequence evolution of the number density 
field of such system is displayed in Fig.~\ref{fig:plot_no_csm}.

%%%%%%%%%%%%%%%%%%%%
% PWN/SN interaction and evolution
%%%%%%%%%%%%%%%%%%%%
The pulsar wind is set in the dense supernova ejecta which has expanded up to a few $\rm pc$ into the local ISM 
(Fig.~\ref{fig:plot_no_csm}a). About $2.5\, \rm kyr$ later, the PWN has adopted a structure made of the 
freely-expanding pulsar wind, a PWN termination shock, a PWN/ejecta inner contact discontinuity (black contour) 
and a transmitted PWN forward shock propagating into the unshocked supernovae ejecta. 
%%%%%%%%%%%%%%%%%%%%
% Ejecta / ISM interaction
%%%%%%%%%%%%%%%%%%%%
The ejecta-ISM interaction leads to the formation of an outer ejecta-ISM contact discontinuity  
affected by strong Richtmyer-Meshkov instabilities~\citep{kane_apj_511_1999}, while 
a reflected supernova termination shock forms and propagates inwards heating the pulsar wind. 
Meanwhile, the supernova forward shock propagates through the ISM (Fig.~\ref{fig:plot_no_csm}b).

The system keeps on growing in size over the next few $\rm kyr$, revealing an efficient mixing of 
pulsar and ejecta materials. The magnetised pulsar wind develops a  bipolar jet along its 
rotation axis, see black contours of Fig.~\ref{fig:plot_no_csm}c.
At time $7.5\, \rm kyr$ the PWN harbors an equatorial disc-like feature, normal to the 
bipolar jet which has gone through the unstable region where the mixing of materials 
takes place and it penetrates the outer layer of shocked ISM gas of the SNR 
(Fig.~\ref{fig:plot_no_csm}d).  
The magnetised PWN, at this point, displays the typical cross-like anisotropic morphology 
described in~\citet{komissarov_mnras_349_2004}, embedded by an overall spherically-symmetric 
expanding blastwave (Fig.~\ref{fig:plot_no_csm}e). 
Last, the jet of the pulsar catches up the supernova remnant shock wave and begins to 
expand into the ISM, generating polar bow shocks (Fig.~\ref{fig:plot_no_csm}f).

%%%%%%%
% Main result form this reference run
%%%%%%%
%
During the evolution, the supernova ejecta high speed compensates the ISM ram pressure, 
allowing the SNR, and, consequently, the PWN morphology to conserve an isotropic symmetry.

%%%%%%%%%%%%%%%%%%%%%%%%%%%%%%%%%%%%%%%%%%
%%%%%%%%%%%%%%%%%%%%%%%%%%%%%%%%%%%%%%%%%%
\subsection{Effects of the progenitor's wind-ISM interaction}
\label{result_2}
%%%%%%%%%%%%%%%%%%%%%%%%%%%%%%%%%%%%%%%%%%%
%%%%%%%%%%%%%%%%%%%%%%%%%%%%%%%%%%%%%%%%%%%
%
In the second model, all evolution stages of the massive progenitor are considered before the 
supernova explosion occurs and the pulsar starts blowing into the former stellar 
wind (Fig.~\ref{fig:plot_csm}). 
The gas distribution at the onset of the pulsar wind is complex, as the wind-ISM interaction 
shaped it throughout the previous star's life. Its arc-like nebula is composed of the large-scale 
component produced by the main-sequence stellar wind plus the shell of red supergiant wind that has 
been subsequently released in it~\citep{meyer_2014bb}. 
The pre-pulsar CSM is therefore a central region of expanding ejecta surrounded by a low-density cavity 
opposite of, with a high-density bow shock facing  the direction of the progenitor's motion 
(Fig.~\ref{fig:plot_csm}a), see also~\citet{meyer_mnras_450_2015,meyer_mnras_502_2021}. 
The asymmetric propagation of the supernova blastwave is strongly constrained by the walls of the cavity. 
It is channelled by the cavity in the direction opposite of that of the stellar motion 
shock has gone through the CSM (Fig.~\ref{fig:plot_csm}b).

The global organisation of the SNR and its central PWN exhibits qualitative dissimilarities 
to that of Fig.~\ref{fig:plot_no_csm} as the game of shock reflections 
generates a more complex object. The morphology of the pulsar wind-ejecta interface 
is governed by that of the ejecta region, adopting its ovoidal shape 
(Fig.~\ref{fig:plot_csm}c) different to that of the model without the progenitor's 
stellar wind (Fig.~\ref{fig:plot_no_csm}c). 
At the time $7.5\, \rm kyr$, the supernova ejecta are channelled into the tubular 
region of shocked stellar wind generated by the progenitor's motion and the pulsar 
wind develops a bipolar jet, although the pulsar wind remains equatorially asymmetric
(Fig.~\ref{fig:plot_csm}d). 
The nebula further evolves such that the inner contact discontinuity of the PWN 
shapes as a function of (i) the pulsar magneto-rotational properties and (ii) the 
supernova shock wave material that is reflected towards the center of the explosion. 
Finally, the contact discontinuity recovers the overall morphology of a bipolar jet 
normal to an equatorial structure (Fig.~\ref{fig:plot_csm}e), with important 
persisting equatorial dissymmetries (Fig.~\ref{fig:plot_csm}f).

%%%%%%%%%%%%%%%%%%%%%%%%%%%%%%%%%%%%%%%%%%%%%%%%
%%%%%%%%%%%%%%%%%%%%%%%%%%%%%%%%%%%%%%%%%%%%%%%%
%%%%%%%%%%%%%%%%%%%%%%%%%%%%%%%%%%%%%%%%%%%%%%%%

\section{Discussion and conclusion}
\label{discussion}

% First time
In this paper, we numerically investigate, in the context of a runaway massive supernova progenitor 
and within the frame of the ideal magneto-hydrodynamics, the influences of the CSM shaped 
during all the progenitor's evolutionary stages on the long-term ($\sim \rm kyr$) morphological development 
of the PWN after the pulsar birth and on the instabilities growing at shocks. 
Other works concerning the release of pulsar winds in SNR were set in the frame of ideal 
hydrodynamics~\citep{Blondin_etal2001ApJ...563..806B,vanderswaluw_aa_397_2003, 
vanderswaluw_aa_404_2003,vanderswaluw_aa_420_2004,Slane_etal2018ApJ...865...86S} as well 
as magneto-hydrodynamics, see~\citet{olmi_mnras_494_2020} and references therein. 
Nevertheless, if some studies tackle 
the problem of runaway pulsars~\citep{olmi_mnras_488_2019}, or even investigated the effects 
of a stratified ISM~\citep{kolb_apj_844_2017}, none of them include in detail the stellar 
wind feedback of the core-collapse progenitor. 
In our scenario, the pulsar wind-supernova ejecta system is embedded into a dense CSM,  
where a large part of the progenitor mass lays, released as pre-supernova stellar wind. 
%}
%
%
%
%
\textcolor{black}{
Note that our models assume that the pulsar is static in the frame of 
reference of the runaway star, so we are neglecting the birth kick 
that many pulsars receive from the supernova explosion. Assuming a 
typical kick velocity of $\approx 400\, \rm km\, \rm s^{-1}$~\citep{verbunt_aa_608_2017}, the pulsar 
would be displaced by about $6\, \rm pc$ over the $15\, \rm pc$ that we 
simulate. Therefore our results apply mainly to the low-velocity
pulsar sub-population ($\le 50\, \rm km\, \rm s^{-1}$), which comprises $\approx 2-5\%$ of all 
pulsars~\citep{igoshev_mnras_494_2020}. 
}

% What we did, message 
Our study shows, in the particular context of a fast-moving 
red supergiant star  in the Galactic plane, that the wind-blown bubble of a 
core-collapse supernova progenitor has a governing impact on the morphology of 
its subsequent PWN. 
As early as $\sim 2.5\, \rm kyr$, the distribution of the contact discontinuity between magnetised 
pulsar wind and supernova ejecta adopts an oblong shape as a result of the anisotropic 
ejecta distribution. 
Since core-collapse SNRs shape according to their CSM, PWNs similarly take their 
morphology as a direct consequence of their progenitor's stellar evolution history. 
Because a significant fraction of massive stars are runaway objects, our findings imply that the stellar 
wind history should not be neglected in the understanding of PWN and that the CSM-induced asymmetries 
should account for up to $10\%$$-$$25\%$ of all individual PWNs.

The stellar wind  history that we use if that of a 
$20\, \rm M_{\odot}$ supergiant star which is amongst the most common \textcolor{black}{progenitors of} core-collapse 
SNRs~\citep{katsuda_apj_863_2018}, although higher-mass evolutionary channels might exist, i.e. 
involving Wolf-Rayet progenitors. 
The bulk motion of the star is taken to be within that  
of the most common runaway stars~\citep{blau1993ASPC...35..207B}. However, 
\textcolor{black}{a small fraction of high-mass stars move with}~\citep{lennon_aa_619_2018} 
and might form PWNs for which the influence of the CSM is milder. 
\textcolor{black}{
Given that the morphology of PWNs retains information} from the stellar evolution history of massive stars, 
our model applies to objects in the Galactic disc region ($\sim 1\, \rm cm^{-3}$) 
rather than in the high-latitude 
parts of the Milky Way, where the low-density medium induce an extended CSM~\citep{meyer_mnras_496_2020} 
with which supernova shock waves (and eventually pulsar winds) weakly interact. 
Conversely, massive stars 
\textcolor{black}{running through dense molecular regions} are more prone to form complex CSM, and, 
therefore, to produce very asymmetric PWNs. 
%

% Observational implications. 
We generate predictive images for our asymmetric PWN from a runaway massive star. 
Fig.~\ref{fig:plot_synchrotron_1} plots non-thermal radio synchrotron emission maps 
using the emissivity $\propto p B_{\perp}^{(s+1)/2}$ with $p$ the thermal pressure of the gas, 
$B_{\perp}$ the component of the magnetic field along the observer's line-of-sight and $s=2$  
the power-law index of the non-thermal electrons distribution~\citep{jun_apj_472_1996}. 
The images are produced with the {\sc radmc-3d} code~\citep{dullemond_2012}, 
displayed with (top) and without CSM (bottom), for selected time instances, and assuming an 
inclination angle of the pulsar is $\theta_{\rm obs}=45\degree$ to the plane of the sky. 
The pulsar wind is at first not visible in the region facing the progenitor's direction of 
motion, where the supernova shock wave interacting with the wind bubble dominates 
(Fig.~\ref{fig:plot_synchrotron_1}a,d). Once the pulsar polar jet grows and penetrates 
the dense region of stellar wind and ejecta, it becomes brighter than that in the 
cavity (Fig.~\ref{fig:plot_synchrotron_1}b,c). This projected up-down surface brightness 
asymmetry increases with time, as the front jet starts to interact with the shocked supernova 
material and the other-side jet continues to propagate in the rarefied medium of the stellar wind cavity.

% Motion pulsar
\textcolor{black}{
%The wind and ISM materials of the surroundings of massive stars is placed into an HII 
%region that is filled with ionizing radiation and which heats the CSM. 
%
Our results show that the effects of the progenitor's stellar wind impacts the 
global morphology of PWNs, and, consequently it  influences the properties 
of the shocks therein. 
}
This should significantly modify the injection and acceleration physics of the relativistic particle 
population, which, in their turn, are responsible for the non-thermal emission properties 
of the PWNs.

We intend to extend this work to a broader study investigating the parameter space of 
various massive stellar progenitors, relativistically treating the pulsar wind and including 
the kick received by pulsar.
%
%It will affect the observation at high energy such as the TeV band~\citep{Hess_aa_612_2018}.

% To finish observation
% We could delete this phrase...
%Last, let mention that this study constitutes a preparatory work for the forthcoming 
%{\it Cherenkov Telescope Array  (CTA)} which will observe supernova remnants as priority 
%targets at $\gamma$-rays and very high-energies and provide insights into the detailed 
%organisation of pulsar wind nebulae.

%%%%%%%%%%%%%%%%%%%%%%%%%%%%%%%%%%%%%%%%%%%%%%%%%%%%%%%%%%%%%%%%%%%%%%%%%%%%%%%%%%%%%%%%%%%
%%%%%%%%%%%%%%%%%%%%%%%%%%%%%%%%%%%%%%%%%%%%%%%%%%%%%%%%%%%%%%%%%%%%%%%%%%%%%%%%%%%%%%%%%%%
%%%%%%%%%%%%%%%%%%%%%%%%%%%%%%%%%%%%%%%%%%%%%%%%%%%%%%%%%%%%%%%%%%%%%%%%%%%%%%%%%%%%%%%%%%%

\section*{Acknowledgements}
\textcolor{black}{
The authors thank the referee, W.~Henney, for advice which improved 
the quality of the paper. 
}
The authors acknowledge the North-German Supercomputing Alliance (HLRN) for providing 
HPC resources that have contributed to the research results reported in this paper.

\section*{Data availability}
%
%This research made use of the {\sc pluto} code developed at the University of Torino  
%by A.~Mignone (http://plutocode.ph.unito.it/) 
%and of the {\sc radmc-3d} code developed at the University of Heidelberg by C.~Dullemond 
%(https://www.ita.uni-heidelberg.de/$\sim$dullemond/software/radmc-3d/).
%
%The figures have been produced using the 
%Matplotlib plotting library for the Python programming language (https://matplotlib.org/). 
The data underlying this article will be shared on reasonable request to the corresponding 
author.

%%%%%%%%%%%%%%%%%%%%%%%%%%%%%%%%%%%%%%%%%%%%%%%%%%%%%%%%%%%%%%%%%%%%%%%%%%%%%%%%%%%%%%%%%%%
%%%%%%%%%%%%%%%%%%%%%%%%%%%%%%%%%%%%%%%%%%%%%%%%%%%%%%%%%%%%%%%%%%%%%%%%%%%%%%%%%%%%%%%%%%%
%%%%%%%%%%%%%%%%%%%%%%%%%%%%%%%%%%%%%%%%%%%%%%%%%%%%%%%%%%%%%%%%%%%%%%%%%%%%%%%%%%%%%%%%%%%

\bibliographystyle{mnras}
\bibliography{grid} % if your bibtex file is called example.bib

\bsp	% typesetting comment
\label{lastpage}
\end{document}